\documentclass[twocolumn,showpacs,amsmath,amssymb,prb,superscriptaddress]{revtex4}

\usepackage{graphicx}
\usepackage{bm}

\begin{document}

\title{Multiple Vortex Cores in 2D Electronic Systems with Proximity Induced Superconductivity}

\author{ N.B.\ Kopnin } \affiliation{ O.V. Lounasmaa Laboratory,
Aalto University, P.O. Box 15100, 00076
Aalto, Finland} \affiliation{ L.~D.~Landau Institute for
Theoretical Physics, 117940 Moscow, Russia}
\author{ I.M.\ Khaymovich}
\affiliation{Institute for Physics of Microstructures, Russian
Academy of Sciences, 603950 Nizhny Novgorod, GSP-105, Russia }
\author{A.S.\ Mel'nikov}
\affiliation{Institute for Physics of Microstructures, Russian
Academy of Sciences, 603950 Nizhny Novgorod, GSP-105, Russia }

\date{\today}

\begin{abstract}
The structure of a proximity induced vortex core in a two-dimensional (2D)
metallic layer covering a superconducting half-space is calculated.
We predict formation of a multiple
vortex core characterized by two-scale behavior of the local density of states (LDOS).
For coherent tunneling between the 2D layer
and the bulk superconductor, the spectrum has two subgap branches while for incoherent tunneling only one of them remains. The
resulting splitting of the zero-bias anomaly and the multiple
 peak structure in the  LDOS should be visible in the tunneling spectroscopy
experiments.
\end{abstract}

\maketitle

Experimental study of subgap quasiparticle states in a
superconductor (SC) placed in a magnetic field provides a
unique tool for probing the internal structure of Cooper pairs.
These states bound to the vortex cores are strongly affected by
the superconducting gap anisotropy in the quasimomentum space (see [\onlinecite{AnisDeltaVortex}] and refs. therein)
and by the number of the order parameter components \cite{Koshelev,Giubileo01}. Direct
information of the spectrum and of the wave functions of such
excitations can be obtained by scanning tunneling
microscopy/spectroscopy (STM/STS) 
 which probes the energy and
spatial dependencies of the local density of states
(LDOS) \cite{oldSTM}. However, the existing experimental data often
provide rather controversial information on the so-called zero
bias anomaly known to be a fingerprint of the Caroli--de
Gennes--Matricon (CdGM) states within the vortex core \cite{CdGM}.
An obvious reason for such ambiguity can be a defect surface layer
which masks the bulk quasiparticle states. For example, an energy
gap in such (possibly nonsuperconducting) layer can appear due to
proximity to the superconductor (see, e.g.,
[\onlinecite{McMillan}]). The vortex states in the systems with
proximity induced superconductivity have been recently studied
using various phenomenological approaches aimed to describe the
hybrid structures consisting of graphene layers coupled to
superconducting electrodes
 \cite{Graphene-SC-1,Graphene-SC-2,Graphene-SC-3,Graphene-SC-our}.

Here we propose the microscopic description of
the vortex core states in two-dimensional (2D) electronic systems with proximity
induced superconductivity and analyze the masking effects of a thin
surface layer on the STM/STS data. Based on the general
approach developed in Ref.~\cite{KopninMelnikov11} for proximity
induced superconductivity we formulate two models which
describe the electron transfer between a 2D
system and a bulk superconductor: (i) coherent momentum-conserving tunneling model
and (ii) incoherent tunneling model that accounts for disorder and
corresponding breakdown of momentum conservation for tunneling
quasiparticles. Within both these models, proximity to a superconductor
induces superconducting correlations in the 2D layer and leads to formation of an energy gap $\Delta_{2D}$ with a magnitude depending
on the tunneling rate $\Gamma$ \cite{KopninMelnikov11,AVolkov95,Fagas-etal-05}:
$\Delta_{2D}\approx \Gamma$ for $\Gamma\ll\Delta$, where $\Delta$
is the gap in the superconducting electrode.
The hallmark of the induced gap is that it does not have a separate critical
temperature but rather vanishes together with the bulk gap.
Quite naturally, the spatial behavior of quasiparticle wave
functions in the 2D layer is determined by two length scales: (i)
the coherence length, $\xi_S = \hbar V_F/\Delta$ for clean or $\xi_S = \sqrt{\hbar D_S/\Delta} $ for dirty limit, characterizing
the bulk electrode and (ii) the 2D coherence
length, $\xi_{2D} = \hbar V_{2D}/\Delta_{2D}$ or $\xi_{2D} = \sqrt{\hbar D_{2D}/\Delta_{2D}}$, where $V_F$, $V_{2D}$
 and $D_S $, $D_{2D} $ are the Fermi velocities and diffusion constants in the bulk and in the 2D layer, respectively. Since $\Delta_{2D}<\Delta$ the coherence length $\xi_{2D}$ usually is longer than $\xi_{S}$.

We show that the proximity induced vortex in a ballistic 2D layer has a
``multiple core'' structure characterized by the two length scales, $\xi_S$ and $\xi_{2D}$. Such a two-scale feature did not appear in the preceding theoretical works where proximity vortex states have been induced by
a primary vortex pinned at a large-size hollow cylinder \cite{Rakhmanov11,Iosel_Feig}.
We calculate the energy spectrum of core excitations for both coherent and incoherent tunneling. For coherent tunneling, the spectrum of
quasiparticles bound to the multiple core consists of two
anomalous branches as functions of the impact parameter $b$. One
branch, $\varepsilon_1 (b)$, qualitatively follows the
usual CdGM anomalous spectrum $\varepsilon_0 (b)$ of the primary
vortex; it extends above the induced gap where it turns into a
scattering resonance. The other branch, $\varepsilon_2
(b)$, lies below the induced gap and resembles the
CdGM anomalous spectrum for a vortex with a much larger core radius
$\sim\xi_{2D}$. This branch has a much slower dependence on the impact parameter and
reaches the induced gap for trajectories that completely
miss the core of the primary vortex.

We demonstrate that the structure of the multiple core is strongly affected by
disorder, i.e., by impurity scattering inside the
bulk electrode or inside the 2D layer, as well as by the barrier disorder.
In our incoherent tunneling model, the latter is accounted for by ensemble averaging over various realizations of disorder in the barrier,
which results in the suppression of influence of the primary CdGM spectral
branch on the spectral characteristics of the 2D layer.
The lower anomalous branch $\varepsilon_2 (b)$ survives the
destructive influence of the barrier disorder, though being shifted
and broadened due to the momentum uncertainty. The impurity
scattering inside the bulk superconductor and inside the 2D layer causes
further smearing of the spectral characteristics of the core states which approach the usual LDOS for dirty superconductors scaled with the corresponding coherence lengths $\xi_{2D}$. As a result, the spatial and energy
dependence  of the LDOS inside the multiple core reveals a rich
behavior dependent on the above spectral properties. The LDOS
contribution from the larger core region $\sim \xi_{2D}$ behaves
similarly to the standard vortex LDOS with the corresponding gap
and coherence length. Effects of primary vortex core on the spatial LDOS
pattern in the 2D layer can be seen as a narrow peak which strongly
depends on the degree of disorder.

\paragraph*{ Model.}
Hereafter we consider a superconducting half space coupled by
quasiparticle tunneling to a 2D covering normal metal layer.
 We
start from the quasiclassical Eilenberger equations for the
retarded or advanced Green function in the 2D layer which are
easily derived from the results of Ref.~\cite{KopninMelnikov11}
(see Appendix~\ref{app1-Eilen-deriv} for details):
\begin{eqnarray} 
-i\hbar{\bf v}_{2D} {\bm \nabla}\check g({\bf p}_{2D},{\bf r}) -
\epsilon \left[ \check \tau_3 \check g({\bf p}_{2D} ,{\bf
r})-\check g({\bf p}_{2D} ,{\bf r})\check \tau_3
\right]\nonumber \\
-\left[\check \Sigma_T \check g({\bf p}_{2D},{\bf r}) - \check
g({\bf p}_{2D},{\bf r})\check \Sigma_T \right]=0 \
,\label{eq-Eliashb-st}
\end{eqnarray}
where ${\bf p}_{2D}$ and ${\bf v}_{2D}=\partial\epsilon_{2D}({\bf
p})/\partial{\bf p}$ are the 2D layer Fermi momentum and velocity.
The Pauli matrices $\tau_1$, $\tau_2$, $\tau_3$, the Green function
\[
\check g =\left(\begin{array}{cc} g & f\\ -f^\dagger & -g
\end{array}\right)\ ,
\]
and the self energy are matrices in the Nambu space.

Coherent tunneling conserves in-plane momentum ${\bf
p}_{2D}$. Therefore, the self energy $\check \Sigma_T$ has the form
\begin{equation}
\check \Sigma_T ({\bf p}_{2D},{\bf r}) = \frac{i\Gamma}{2} \left[
\check g_S ({\bf p}_{+};{\bf r}, 0) + \check g_S ({\bf
p}_{-};{\bf r}, 0)\right], \label{selfen-coh}
\end{equation}
where 3D momentum ${\bf p}_\pm = ({\bf p}_{2D}, \pm p_z)$ lies on the Fermi
surface of the bulk SC, $p^2_{2D}+p^2_{z}=p^2_F$.
 Coherent tunneling is impossible if the Fermi momentum in the 2D layer
 is larger than that in 3D, $p_{2D}>p_F$. For incoherent tunneling,
 the in-plane momentum is not conserved. All momentum directions participate in tunneling, thus
\begin{equation}
\check \Sigma _T({\bf r})=i\Gamma \left< \check g_S({\bf p}_F;{\bf
r},0)\right> \label{selfen} \ .
\end{equation}
The angular brackets denote the averaging over directions of the 3D
Fermi momentum ${\bf p}_F$. The tunneling rate $\Gamma
\sim t^2/E_{2F}$ can be expressed \cite{KopninMelnikov11}
 in terms of the normal-state tunnel conductance $G=1/RS$ per unit contact area,
$
\Gamma=G/(4\pi G_0\nu_2) \sim E_{2F} R_0 /R \ ,
$
with the conductance quantum $G_0=e^2/\pi \hbar$ and
the normal 2D density of states (DOS) $\nu_2=m/2\pi \hbar^2$.
Therefore $ \Gamma/E_{2F}\ll 1$ if the total tunnel resistance $R$
is much larger than the  Sharvin resistance $R_0=(NG_0)^{-1} $
 for an ideal $N$-mode contact with the contact area $S$ \cite{Datta}.
 Nevertheless, there is a room for the condition
 $\Gamma \sim \Delta$ to be fulfilled even for a large contact resistance $R\gg R_0$.

Here we restrict ourselves to the limit of low tunneling rate $\Gamma \ll \Delta$
 which leads to a small induced gap \cite{KopninMelnikov11} $\Delta_{2D}=\Gamma$ and long coherence length $\xi_{2D}\gg \xi_{S}$.
We consider an isolated vortex line oriented along the $z$ axis perpendicular to
the SC/2D interface and choose the gap function inside the bulk
superconductor in the form $\Delta =\Delta_0(\rho) e^{i\phi}$,
where $(\rho , \phi, z)$ are the cylindrical coordinates;  $\Delta_0(\rho)$ approaches the bulk value
$\Delta_\infty$ far from the vortex core.  The
self energies in the 2D layer are given by
Eqs.~(\ref{selfen-coh}) and (\ref{selfen}). They have parts with sharp peaks localized at small
distances $\rho\sim\xi_S$ and the adiabatic ``vortex potential'' part
$\Delta_{2D}\sim\Gamma  e^{i\phi}$ which defines the large scale behavior
of the 2D layer Green functions.

\paragraph*{Multiple core.\ Clean limit with coherent
tunneling.} To elucidate the basic features of the multi-scale
vortex core in the 2D layer we consider first an idealized picture
without any disorder assuming  specular electron reflection at the
surface of the bulk SC.

For the low-energy limit $\epsilon\ll\Delta_\infty$ one can find the induced
vortex potential at large distances $\rho\gg\xi_S$:
\begin{eqnarray}
\check \Sigma _T=i\Gamma\check g^{R\left( A\right) }_S \simeq
i\Gamma\check\tau_2e^{i\check\tau_3\phi} \ .\label{fg-asimpt}
\end{eqnarray}
The quasiparticles propagating along the trajectories that miss the
primary vortex core ($b>\xi_S$) are affected only by this long-distance ($\xi_{2D}\gg\xi_S$) part of the induced gap potential and the corresponding
solutions for the Green functions coincide with the standard CdGM
ones for the gap value replaced with $\Gamma$.
A quasiclassical trajectory can be
parameterized by its angle $\alpha$ with the $x$ axis, the impact
parameter $b=\rho \sin (\phi -\alpha )$ and the coordinate $s=\rho
\cos (\phi -\alpha )$ along the trajectory. The corresponding
anomalous spectrum for 2D excitations is \cite{KramerPesch,Kopnin-book}
\begin{equation}
\epsilon=\epsilon_2(b)=\frac{2\Gamma^2 b}{\hbar V_{2D}} \ln
\Lambda \ ,\label{eps2}
\end{equation}
where $\Lambda = \xi_{2D}/|b|$.
This modified CdGM branch should dominate in the local
DOS at large distances $\rho\gg\xi_S$.

Trajectory with a small impact parameter $b\lesssim \xi_S$ can be divided into the long-distance part going far from the primary vortex core, and the region inside the core. Far from the core, the solution is found using the vortex potentials Eq.~(\ref{fg-asimpt}).
In the core region one should take into
account the self-energy parts localized within the
primary vortex \cite{KramerPesch,Kopnin-book}. We put ${\check \Sigma_T= i\Gamma
\left({\bf F}_0 \cdot\check {\bm \tau}\right)}$, where $
\check{\bm \tau} =(\check\tau_1
e^{-i\check\tau_3\alpha},\check\tau_2
e^{-i\check\tau_3\alpha},\check\tau_3)$,
the vector ${\bf F}_0=(-\zeta_S,\theta_S,g_S)$ is normalized by  ${\bf F}_0^2=1$ and
has the components
\begin{eqnarray}
\zeta_S  =\frac{\hbar v_{\parallel }e^{-K}}{2Q\left[ \epsilon -\epsilon
_{0}\pm i\delta \right]} , \; \theta_S  = \frac{2}{\hbar v_{\parallel
}}\int_{0}^{s}(\epsilon - \frac{b\Delta_0}{\rho^\prime}
)\zeta_S ds^\prime ,   \label{zeta}\\
\epsilon _{0}(b) =b Q^{-1}\int_{0}^{\infty }[\Delta _{0}/\rho ]%
e^{-K(s)}\,ds \ ,  \label{Eb}\\
Q=\int_{0}^{\infty }e^{-K(s)}\,ds \ ;\quad K(s)=
\frac{2}{\hbar v_{\parallel}}\int_{|b|}^{\rho}\Delta
_{0}(\rho^\prime) \,d\rho^{\prime }
  \label{K} \ .
\end{eqnarray}
Here ${\bf v}_\parallel$ is the 3D Fermi velocity projection onto the $(x,y)$ plane.
The upper (lower) sign of an infinitely small $\delta>0$ refers to the retarded (advanced) function.
\begin{figure}[t]
\includegraphics[width=0.80\linewidth]{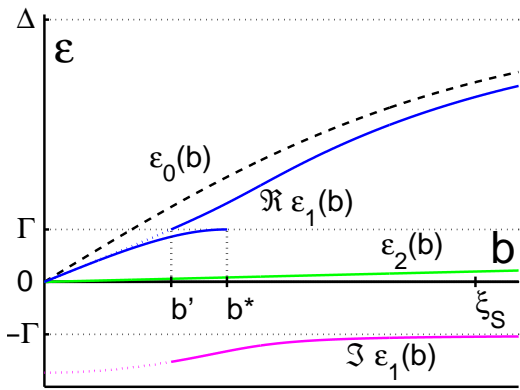}
\caption{(Color online) Two-scale behavior of the spectrum, Eq.~(\protect\ref{pole_coh}), for coherent tunneling.
The spectrum has two localized branches, $\epsilon_1(b)$ and $\epsilon_2(b)$, for $\epsilon<\Gamma$. The branch $\epsilon_2(b)$ has a scale $\xi_{2D}$, it saturates at $\epsilon = \Gamma$ for $b\gg \xi_{2D}$.
$\epsilon_1(b)$ has a scale $\xi_S$. For $\Re\epsilon >\Gamma$ it transforms into scattering
resonances. $b^*$ is defined as $\epsilon_1(b^*)=\Gamma-0$, while
$b'$ corresponds to $\Re\epsilon_1(b')=\Gamma+0$.} \label{fig-spectrum2}
\end{figure}
Matching the 2D Green functions through the primary core region of
rapidly changing self-energy potentials (see Appendix
\ref{app3-scale-separation-solution}) we find both the spectrum
\begin{eqnarray}
q^{-1}[\epsilon-\epsilon_2(b)][\epsilon -\epsilon_0(b)] + \Gamma^2 - \Gamma \sqrt{\Gamma^2-[\epsilon-\epsilon_2(b)]^2} =0 \label{pole_coh}
\end{eqnarray}
and the Green functions for trajectories with $b \ll \xi_{2D}$. Here $q= v_\parallel/V_{2D}$. For $b \lesssim \xi_S$, the cut-off parameter in Eq.\ (\ref{eps2})  should be replaced with
$\Lambda=\xi_{2D}/\xi_S$.
\begin{figure}[t]
\includegraphics[width=0.8\linewidth]{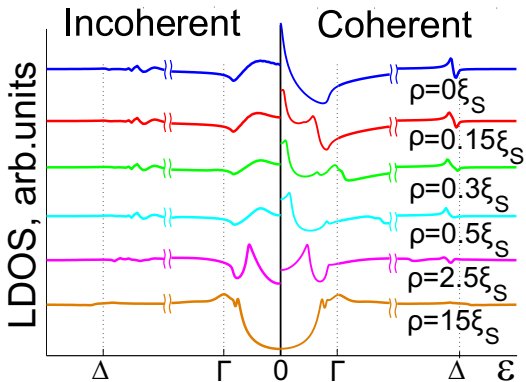}
\caption{(Color online) The local DOS in logarithmic scale for coherent
(right panel) and incoherent (left panel) tunneling in the
clean limit for different distances $\rho$ from the vortex center. The peaks in LDOS exist up to distances $\sim \xi_{2D}$.
Here $\Delta/\Gamma=5$, $q=1$. }
\label{fig-DOS}
\end{figure}

The resulting two-scale behavior of the spectral branches is
illustrated in Fig.~\ref{fig-spectrum2}.  The complex-valued
energy branches satisfy the symmetry condition:
 $\epsilon_{1,2}(-b)=-\epsilon^*_{1,2}(b)$. There are two real-valued energy
branches in the range $|\epsilon|<\Gamma$ crossing zero of energy as functions of the
impact parameter and one complex-valued branch in the range $\Gamma<|\epsilon|<\Delta_\infty$. The
lowest-energy branch  $\epsilon_2(b)$ as a function of the impact parameter has a characteristic scale $\xi_{2D}$: For
$b\lesssim \xi_{2D}$ it is determined by Eq.~(\ref{eps2}) with the proper cut-off parameter $\Lambda$ as discussed above. On the other hand, it saturates at $\epsilon=\Gamma$ for
$b>>\xi_{2D}$. The branch $\epsilon_1(b)$ has a scale $\xi_S$:
At low energies it goes slightly below the CdGM spectrum $\epsilon_0(b)$
in the bulk SC,
 $\epsilon_1(b) =(1+q/2)^{-1}\epsilon_0(b)$.
Above the induced gap $\Gamma$ the spectrum
transforms into a scattering resonance due to the decay into the
delocalized modes propagating in the 2D layer: $\epsilon_1(b)
=\epsilon_0(b)-i\Gamma$ for $|\epsilon|\gg\Gamma$.
Since Eq.~(\ref{pole_coh}) determines a pole of the retarded
Green function in the lower half-plane of complex $\epsilon$,
the square root in Eq.~(\ref{pole_coh}) should be analytically
continued under the cut extending from $-\infty$ to $-\Gamma$ and
from $\Gamma$ to $+\infty$. As a result, $\epsilon_1(b)$ has a
discontinuity at $\epsilon_1=\Gamma$.

Two branches appear because the system under
consideration consists of two sub-systems \cite{Shiozaki12}, the
bulk SC and the 2D proximity layer, each with its own anomalous
branch. The branch $\epsilon_1(b)$ is the proximity image of the
bulk spectrum $\epsilon_0(b)$ with the spectral weight
proportional to the tunneling probability $\Gamma$. The branch
$\epsilon_2(b)$ belongs to the 2D layer itself. We note that the
presence of two anomalous branches does not contradict to the
index theorem \cite{Volovik}. Indeed, its application requires that
both zero of the quasiclassical Hamiltonian at the Fermi surface
and its singularity at $\epsilon =\epsilon_0(b)$ are taken into
account when calculating the topological invariant. As a result,
the number of anomalous branches is increased up to 2 for a
single-quantum vortex.

The multiple-branch spectrum results in multiple peaks in the
 LDOS energy dependence (right panel in Fig.~\ref{fig-DOS}). 
 Here the local DOS is defined by the angle-resolved one (normalized by its normal state value)
${N_\epsilon(s,b)= [g^R(s,b)-g^A(s,b)]/2}$ averaged over the trajectory direction.
The multiple peak structure appears to be most pronounced deeply
inside the primary core region (at distances $\rho \lesssim
\xi_S^2/\xi_{2D}$ when $\epsilon_1<\Gamma$) illustrating, thus,
the two-scale structure of the vortex core.

The number of LDOS
peaks at a certain distance $\rho$ from the vortex center is
determined by the number of spectral branches at $b\sim \rho$. The
spectrum discontinuity at the induced gap $\Gamma$ causes the
appearance of three LDOS peaks
 in the range of distances, corresponding to
$b'<b<b^*$ (see the plot for $\rho=0.3\xi_S$ in
Fig.~\ref{fig-DOS}). The numerical LDOS patterns have been
obtained by the subsequent solving of two sets of Eilenberger
equations in Riccati parametrization \cite{Schopohl}: first, we
calculated the Green functions for the bulk superconductor with
the model order parameter
 profile $\Delta_0(\rho)=\Delta_\infty\rho/\sqrt{\rho^2+\xi_S^2}$
 and, second, we have found the solution of Eq.~\eqref{eq-Eliashb-st} for a 2D layer with the
 induced potentials defined by the Eq.~\eqref{selfen-coh}.

\paragraph*{Multiple core.\ Clean limit with incoherent tunneling.}
We proceed our study with the consideration of disorder effects
and introduce first the momentum scattering during the tunneling
process described within the incoherent tunneling model.
Considering the tunneling as a perturbation  one can assume a
specular quasiparticle scattering at the interface  and, thus, use
the results of the previous section for the Green functions. The
self-energy potentials in this case can be obtained by averaging
of Eqs.\ (\ref{zeta}-\ref{K}) over the trajectory direction:
$\check\Sigma_T = i\Gamma\left<\check g_S\right>$.
This averaging does not affect,
of course, the induced gap function (\ref{fg-asimpt})
 outside the primary vortex
core and, thus, the spectrum $\epsilon_{2}$ survives the
influence of the tunnel barrier disorder at least for $b>\xi_S$.
On the contrary, the subgap branches localized within the primary
vortex core region appear to be completely destroyed. Such
dramatic consequence of the momentum scattering is caused by the
averaging of electronic wave functions with different impact
parameters and consequent loss of
 any information about the CdGM states of the primary vortex.
A natural consequence of the momentum scattering is the
appearance of a finite broadening of energy levels for
trajectories with small impact parameters $b\lesssim\xi_S$.

Following again the matching procedure described above (see Appendix
\ref{app3-scale-separation-solution})
 we find the angle-resolved DOS for $b\lesssim\xi_S$
and $|\epsilon|\ll\Gamma$,
\begin{eqnarray}
N_\epsilon(s,b)&=& \frac{\Gamma \gamma(b)e^{-|s|/\xi_{2D}}}{[\epsilon-\epsilon_2(b)- \beta(b)]^2
+\gamma^2(b)} \ ,\label{ARDOS}\\
\beta(b)&= &\Gamma^2  \left< \frac{\pi q
}{Q\Omega}{\rm sign}(\epsilon+\Omega b) \right>_z  \ ,
\label{beta-res}\\
\gamma(b)&=&\Gamma^2   \left<
\frac{q}{Q\Omega}\ln \frac{\Delta_\infty}{|\Omega b
+\epsilon|}\right>_z\ . \label{gamma-res}
\end{eqnarray}
The angular brackets denote averaging over the momentum $p_z$
along the vortex axis in bulk SC, $\Omega =
\partial\epsilon_0/\partial b$. The DOS has a peak of the height $\Gamma/\gamma$ at an energy
${\epsilon=\epsilon_2(b)+\beta(b)}$ shifted from a standard bound
state level (see Appendix~\ref{app3-subsec-Incoh} for details).
This shift results in splitting of the zero-bias anomaly in
LDOS, as is seen from our numerical analysis  (see the left panel in
Fig.~\ref{fig-DOS}). For LDOS calculations we use the numerical
procedure similar to that used for the
coherent limit above with the induced potentials
averaged over the Fermi surface (assumed cylindrical) in the bulk.

\paragraph*{Multiple core.\ Dirty superconductor with clean 2D
layer.} Smearing of the energy dependence of the induced
potentials caused by disorder becomes even stronger
if the bulk SC has short mean free path: $\ell \ll \xi_S$. In
dirty limit, the momentum averaged retarded (advanced) Green
functions are parameterized as follows:
\begin{equation}\label{selfen-dirty}
\check g_S^{R(A)}(\rho)=\check \tau_3\sin\Theta^{R(A)}+\check
\tau_2\cos\Theta^{R(A)} e^{-i\check \tau_3\phi} \ .
\end{equation}
We put $\Theta^{R(A)}=\pm \Theta_1 +i \Theta_2$. The boundary conditions \eqref{fg-asimpt} require $\Theta_1 \to \pi/2$, $\Theta_2 \to 0$ for $\rho \to 0$. At large distances $\sin \Theta_1 \to 0$, $\tanh \Theta_2 \to -\epsilon /\Gamma$ for $\epsilon < \Delta_\infty$ while  $\cos \Theta_1 \to 0$, $\tanh \Theta_2 \to - \Delta_\infty/\epsilon $ for $\epsilon > \Delta_\infty$.
Therefore, $\Theta_2=0$ for $\epsilon \ll \Delta_\infty$,
and the Usadel equation becomes \cite{GorkovKopnin}
\begin{equation}
D_S\left[ \nabla^2 \Theta_1 + \frac{\sin (2\Theta_1)}{2\rho^2}  \right] -2\Delta_0\sin \Theta_1 =0 \label{eqUsadel1}
\end{equation}
The solution of Eq.\ (\ref{eqUsadel1})
 has been found in Ref.~\cite{GorkovKopnin}: the function
$\Theta_1(\rho)$  monotonously decays from $\pi /2$  at the origin down to
the zero value at $\rho\gg\xi_S$. The Green functions (\ref{selfen-dirty}) determine the
induced vortex potentials $\check\Sigma_T=i\Gamma\check g_S$.
\begin{figure}[t]
\includegraphics[width=0.7\linewidth]{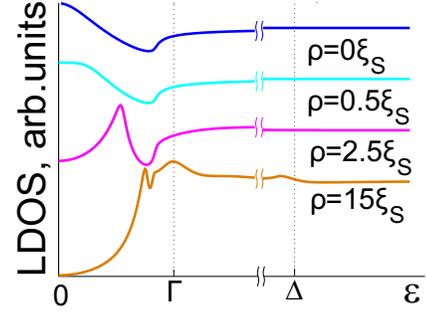}
\caption{(Color online) The local DOS in logarithmic scale for the dirty limit
with the parameters $\Delta/\Gamma=5$, $V_{2D}/V_F=1$ for
different distances $\rho$ from the vortex center. }
\label{fig-DOS-dirty}
\end{figure}
For $|\epsilon| \ll \Gamma$ and $b\lesssim\xi_S$ the peak in the
energy dependence of the angle-resolved DOS is described by the
Eq.\eqref{ARDOS} with $\beta=0$ and
\begin{equation}\label{gamma-res-dirty}
\gamma = \frac{2\Gamma^2}{\hbar V_{2D}} \int_0^{\infty} \sin\Theta_1
\, ds
\end{equation}
(see Appendix~\ref{app3-subsec-Dirty} for details). The numerical
results clearly confirm the existence of one broadened peak in
the LDOS dependence vs energy: this peak shifts with the
increasing distance from the vortex center and becomes sharper
(see Fig.~\ref{fig-DOS-dirty}).
 Our numerical
procedure of the LDOS calculation in this limit is based on the
using of a standard relaxation method \cite{RelaxMethod} for solving the Usadel
equation \cite{Golubov} in the bulk SC and Riccati parametrization for
Eilenberger equations in the 2D layer.

\paragraph*{Vortex core expansion.\ Dirty superconductor and
2D layer.} To complete our analysis we discuss the case of strong
disorder both in the bulk superconductor and in the 2D layer.
This limit has been previously studied in Ref.~\cite{Golubov-vortex}.
As before, one can parameterize the Green functions averaged over the 2D momentum directions in the form of Eq.\ (\ref{selfen-dirty}),
where we use $\Psi$ for the 2D-layer functions instead of $\Theta$. The boundary conditions coincide with those for Eq.\ (\ref{selfen-dirty}) where $\Delta_\infty$ is replaced with $\Gamma$. With the self energies from the previous subsection, the Usadel equation for the retarded function  for
$\epsilon \ll \Delta_\infty$ is
\begin{equation}\label{eqUsadel-2D}
D_{2D}\left[\nabla^2\Psi+\frac{\sin(2\Psi)}{2\rho^{2}}\right]
-2\Gamma\sin(\Psi-\Theta)-2i\epsilon\cos\Psi=0 .
\end{equation}
$\Theta$ is essentially nonzero only inside the primary core
region $\rho<\xi_S$. The condition $\xi_S\ll \xi_{2D}=\sqrt{\hbar
D_{2D}/\Gamma}$ ensures that such short-distance
inhomogeneity in the induced vortex potentials does not disturb the
adiabatic solution based on
Eq.~\eqref{fg-asimpt} (see Appendix~\ref{app3-subsec-Dirty-all}).
Thus, putting $\Theta=0$ in
Eq.~\eqref{eqUsadel-2D} we reduce our problem to that describing a standard vortex in a dirty superconductor with the
gap value $\Gamma$. Thus, the full disordered system should reveal
the same LDOS patterns as in the bulk case, though scaled with the
much larger coherence length $\xi_{2D}$ instead of $\xi_S$.
This conclusion is, of course, in agreement with numerical calculations in
 the Ref.~\cite{Golubov-vortex}.

\paragraph*{Conclusion} To summarize, we calculate the electronic structure of a proximity induced vortex core in a 2D
metallic layer covering a superconducting half-space. We predict
formation of a multiple vortex core resulting in a two-scale
behavior of the LDOS. For coherent tunneling between the 2D layer
and the bulk superconductor, the spectrum has two subgap branches
while for incoherent tunneling only one of them remains. The
splitting of the zero-bias anomaly and the multiple peak structure
in the  LDOS should be visible in the tunneling spectroscopy
experiments. Disorder further smears the multiple peak structure
inside the double-scale vortex core. When both the bulk SC and the
2D layer are in dirty limits, the 2D LDOS qualitatively repeats
that in the bulk SC scaled with the larger coherence length
$\xi_{2D}$. Such expansion of the vortex core probably relates to
the anomalously large vortex images observed in $MgB_2$
 \cite{eskildsen} and high-- $T_c$ cuprates \cite{yeh}.


We thank A.~Buzdin and G.~Volovik for stimulating discussions.
This work was supported in part by the Academy of Finland, Centers
of excellence program 2012--2017, by the Russian Foundation for
Basic Research
, by the Program
``Quantum Physics of Condensed Matter'' of the Russian Academy of
Sciences, and by FTP
``Scientific and educational personnel of innovative Russia in
2009-2013''.

\appendix

\section{Eilenberger equations for coherent and incoherent models}\label{app1-Eilen-deriv}

We start with the equation for the retarded (advanced) Green functions derived in Ref.~\cite{KopninMelnikov11}
\begin{multline}
\check G^{-1}({\bf r}_1)\check G_{2D}({\bf r}_1,{\bf
r}_2,\epsilon)-\!\! \int \check \Sigma _T({\bf r}_1,{\bf r}^\prime
)\check G_{2D}({\bf r}^\prime ,{\bf r}_2,\epsilon)\, d^2 r^\prime \\
=\check 1d^{-1} \delta ({\bf r}_1 -{\bf r}_2)\ .\;
\label{eqG-coord}
\end{multline}
Here $d$ is the layer thickness, $\check \tau_1$, $\check \tau_2$, and $\check \tau_3$ as well as
\[
\check G_{2D}=\left(\begin{array}{cc} G& F\\ -F^\dagger & \widetilde
G\end{array}\right) \ , \;
\check G^{-1}({\bf r}_1)=\epsilon_{2D}(\hat {\bf p})-\mu
-\epsilon\check \tau_3\ ,
\]
are matrices in the Nambu space,
$ \epsilon_{2D}(\hat {\bf p})$ is the spectrum of the 2D electron
system, $\hat {\bf p}=-i\hbar {\bm \nabla}$
and ${\bf r}$ are the 2D momentum and coordinate, correspondingly.
The self-energy takes the form,
\begin{equation}
\check \Sigma _T({\bf r}_1,{\bf r}_2)=   d\, t({\bf r}_1)\check
G_S({\bf r}_1,z_1=0;{\bf r}_2,z_2=0) t({\bf r}_2) \label{sigma}
\end{equation}
where $t({\bf r})$ is the tunneling amplitude which is assumed
real and the Green function $G_S$ of the bulk
 SC is taken at the SC/2D interface $z=0$.
The above equations can be strongly simplified using a standard
quasiclassical procedure which allows us to derive the Eilenberger
equations for quasiclassical Green function
\[
\check g ({\bf
p}_{2D},{\bf r})=({\pi i})^{-1}\int {d\xi_{2}} \check G_{2D} ({\bf
p},{\bf r})
\]
Here we derive expressions for the self energies (\ref{selfen-coh},
\ref{selfen}) and the Eilenberger equations \eqref{eq-Eliashb-st}
for different tunneling models.

\subsection{Coherent tunneling}\label{app1-subsec-Coherent}

Let us assume that the in-plane momentum projection is
conserved during the tunneling process. This amounts for a
tunneling amplitude $t({\bf r})$ independent of the coordinate
along the SC/2D interface. In 2D momentum representation the self
energy in Eq.~(\ref{eqG-coord}) becomes
\begin{eqnarray*}
\check \Sigma _T({\bf p}_1,{\bf p}^\prime)= dt^2\int \check G_S
({\bf p}_1,p_z;{\bf p}^\prime, p_z^\prime)\frac{dp_z\,
dp_z^\prime}{(2\pi)^2} \ .
\end{eqnarray*}
We now apply the operators to the Green function from the right
and subtract this equation from Eq.~\eqref{eqG-coord}. Integrating
the result over the energy variable near
the Fermi surface $\xi_{2}=\epsilon_{2D}({\bf p})-\mu$ and using
\begin{multline*}
\int \frac{d\xi_{2}}{\pi i} \int \check \Sigma _T({\bf p}_1,{\bf
p}^\prime
)\check G_{2D}({\bf p}^\prime ,{\bf p}_2)\frac{d^2 p^\prime}{(2\pi)^2}=
dt^2 g({\bf p}_{2D} ,{\bf r})\\
\times \int\frac{dp_z}{2\pi} \check g_S
({\bf p}_{2D},p_{z};{\bf r}, 0)\, \pi i
\delta_{\Delta}\left[\epsilon_S({\bf p}_{2D},p_z)-\mu\right]
\end{multline*}
we obtain the quasiclassical Eilenberger equation (\ref{eq-Eliashb-st}). The quasiclassical Green function $\check g_S({\bf p};{\bf r},0)$
of the bulk SC is taken at the SC/2D interface
${z=0}$. We use
the mixed momentum ${\bf p}$-coordinate ${\bf r}$
representation describing the relative and center-of-mass motion
of electrons in the Cooper pair and put
\begin{eqnarray*}
\check {\mathcal G}_S ({\bf p},p_z;{\bf r}, z)&=
&\check g_S ({\bf p},p_{z};{\bf r}, z) \pi i \delta_{\Delta}(\xi_3) \ ,\\
\check {\mathcal G}_{2D} ({\bf p} , {\bf r})&=&\check g({\bf p}
,{\bf r}) \pi i \delta_{\Delta}(\xi_{2}) \ .
\end{eqnarray*}
Here $\xi_3=\epsilon_{S}({\bf p},p_z)-\mu$ is the normal quasiparticle
spectrum in the 3D half-space,
$\check g_S $ and $\check g$ are standard quasiclassical Green functions, and
$\delta_{\Delta}(\xi_{2,3})$ is a delta function broadened at the
gap energy scale $\Delta$.

Assuming isotropic Fermi surfaces in both the
superconductor and the 2D layer we get the self energy in the form
of Eq.~\eqref{selfen-coh} with the tunneling rate
$$
\Gamma =  dt^2 \int\limits_0^\infty
\delta_{\Delta}\left[\epsilon_S({\bf p}_{2D},p_z)-\mu\right] dp_z\
.
$$
Provided the 2D Fermi surface is smaller than the extremal cross
section of the 3D Fermi surface, i.e., $p_{2D}<p_F$ the expression
for the tunneling rate reads: ${\Gamma=d m t^2/p_{3z}}$. For large
2D Fermi surfaces $p_{2D}>p_F$ the self energy term vanishes, and
the coherent tunneling is impossible. The case of close momenta
$p_{2D}\simeq p_F$ deserves special consideration which should
take account of a finite delta function width:
 $\Gamma\sim d t^2 (m/\Delta)^{1/2}$.

\subsection{Incoherent tunneling}\label{app1-subsec-Incoherent}

We now  assume that the tunneling occurs through random centers such that the ensemble average of amplitudes in Eq.~(\ref{sigma}) is
\begin{equation}
\left< t({\bf r}_1) t({\bf r}_2)\right> = t^2 s_a \delta({\bf
r}_1-{\bf r}_2 ) \ ,
\end{equation}
where $s_a$ is the correlated area of the order of atomic
scale. After averaging the self energy becomes:
\begin{eqnarray*}
\check \Sigma _T({\bf r}_1,{\bf r}_2)=t^2 ds_a\check G_S({\bf r}_1,{\bf r}_1;0)\delta({\bf r}_1-{\bf r}_2 )\\
= t^2 ds_ai\pi \nu_3(0) \left<\check g_S({\bf p};{\bf
r},0)\right>\delta({\bf r}_1-{\bf r}_2 ) \ .
\end{eqnarray*}
Here $\nu_3(0)$ is the normal density of states in the bulk material.
 Angular brackets denote averaging over
three-dimensional momentum directions. Within the quasiclassical
approach the resulting self energy is given by
Eq.~\eqref{selfen} with the tunneling rate $\Gamma =\pi
\nu_3(0)ds_at^2 $. This approximation coincides with
that used in Ref.~\cite{KopninMelnikov11}.

\section{Induced vortex potentials}\label{app2-IndPot}

In both tunneling models the Green functions of the 2D
layer satisfy the Eilenberger equations \eqref{eq-Eliashb-st}:
\begin{eqnarray}
-i\hbar{\bf v}_{2D}  {\bm \nabla}  f -2\left[\epsilon +\Sigma_1 \right]\! f +2 \Sigma_2 g =0 ,\; \label{El2} \\
i\hbar{\bf v}_{2D}  {\bm \nabla}  f^{\dagger } -2\left[\epsilon +\Sigma_1  \right]\! f^{\dagger} +2\Sigma_2^{\dagger} g =0 ,\;
\label{El3}\\
-i\hbar{\bf v}_{2D}  {\bm \nabla}  g+ \Sigma _2 f ^{\dagger} - \Sigma^\dagger _2 f =0 .\; \label{El1}
\end{eqnarray}
and the normalization condition $g ^2- f  f^{\dagger} =1$ with the
self energy (\ref{selfen-coh}, \ref{selfen}) as effective
potentials
$$\check \Sigma_T= \begin{pmatrix}\Sigma_1 & \Sigma_2\\ -\Sigma^{\dagger}_2 & -\Sigma_1\end{pmatrix} \ .$$

\begin{figure}[t]
\includegraphics[width=0.75\linewidth]{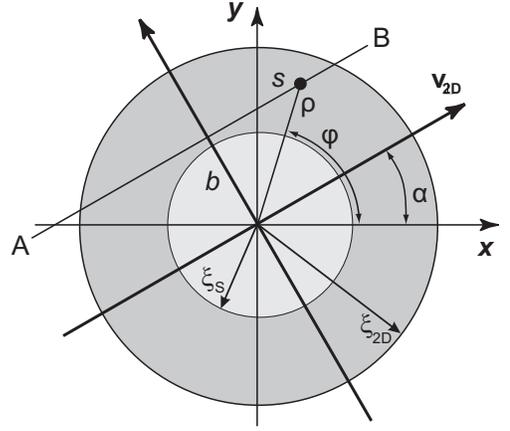}
\caption{(Color online) 
  The coordinate frame near the
multiple vortex core. Primary (induced) core is shown by the white (gray) circle. The quasiparticle trajectory
with an impact parameter $b$ (line AB) passes through the point
$(\rho ,\phi )$ shown by the black dot.}
\label{fig-2Dvortex-coords}
\end{figure}

Quasiparticles are conveniently described by the coordinates along their trajectories (see Fig.~\ref{fig-2Dvortex-coords}). A quasiclassical trajectory is parameterized by its angle $\alpha$ with the $x$ axis, the impact parameter $b=\rho \sin (\phi -\alpha )$ and the coordinate $s=\rho
\cos (\phi -\alpha )$ along the trajectory. We introduce the
symmetric and antisymmetric parts of the Green functions
 \cite{KramerPesch,Kopnin-book}:
\begin{subequations}\label{f-param}
\begin{align}
f=-\left[\zeta (s)+i\theta (s)\right]\exp (i\alpha)\\
f^{\dagger }=\left[\zeta (s)-i\theta (s)\right]\exp (-i\alpha ),
\end{align}
\end{subequations}
where $\zeta (s)=\zeta (-s)$, and $\theta (s)=-\theta (-s)$. The
normalization condition requires $g^{2}+\theta ^{2}+\zeta ^{2}=1$.
Eilenberger equations (\ref{El2}-\ref{El1}) can be rewritten as
follows:
\begin{eqnarray}
\hbar V_{2D} \frac{d\zeta}{ds} +2\left(\epsilon +\Sigma_1\right) \theta  -2ig\Sigma_R &=0  ,\quad \label{eq-zeta-2}\\
\hbar V_{2D} \frac{d\theta }{ds} -2\left(\epsilon + \Sigma_1\right) \zeta  -2ig \Sigma_I &=0 , \quad \label{eq-theta-2}\\
\hbar V_{2D}\frac{dg}{ds} +2i\zeta\Sigma_R+2i\theta\Sigma_I &=0 , \quad  \label{eq-g-2}
\end{eqnarray}
where
\begin{subequations}\label{Sigma_RI}
\begin{align}
2\Sigma_R=\left( \Sigma_2  e^{-i\alpha} +  \Sigma_2^{\dagger}e^{i\alpha} \right) ,\quad\\
2i\Sigma_I=\left( \Sigma_2  e^{-i\alpha} -  \Sigma_2^{\dagger}e^{i\alpha} \right) .\quad
\end{align}
\end{subequations}

In order to evaluate the induced vortex potentials $\Sigma_1$,
$\Sigma_2$ and $\Sigma_2^\dagger$ we start from two important
assumptions: (i) low interface barrier transparency and (ii)
negligible effect of the diffusive interface reflection. These
assumptions allow us to neglect the effect of tunneling on the
bulk superconductor characteristics and use the bulk values of the
quasiclassical Green functions. Restricting our consideration to
the small energy  values $\epsilon\ll\Delta_\infty$ we find the
large-scale ($\rho\gg\xi_S$) self energy \eqref{fg-asimpt} to be
independent of the particular tunneling model and disorder rate
in the bulk SC: $\Sigma_1=i\Gamma g_S\approx 0$, $\Sigma_2=i\Gamma
f_S\approx \Gamma e^{i\phi}$, i.e. $\Sigma_R\approx \Gamma
s/\rho$, $\Sigma_I\approx\Gamma b/\rho$.

Contrary to the large distance limit the induced vortex potentials
close to the primary  vortex core reveal a very peculiar behavior
depending on impurity concentration and momentum conservation
during the tunneling process.
In a clean limit of the bulk SC we
use the Green function parametrization similar to (\ref{f-param})
and rewrite the Eilenberger equations in the following form:
\begin{eqnarray}
\hbar v_{\parallel }{%
\frac{{\partial \zeta_S }}{{\partial s}}}+2\epsilon \theta_S -2i\Delta _{0}g_S s/\rho &=&0,  \label{s} \\
\hbar v_{\parallel }{\frac{{\partial \theta_S }}{{\partial
s}}}-2\epsilon \zeta_S -2i\Delta _{0}g_S b/\rho  &=&0.  \label{a}
\end{eqnarray}

For energies  $\epsilon\ll \Delta_\infty$, the functions $g_S$ and $f_S$,
$f^{\dagger }_S$ are large near the vortex. Assuming that $\zeta_S
^{2}\gg \theta_S ^{2}-1$, we have $g_S^{R(A)}=i\zeta_S^{R(A)}$.
The plus sign here is chosen to satisfy the condition of vanishing
$g_S$ at large distances according to Eq. (\ref{fg-asimpt}). The
solution of Eqs.\ (\ref{s}, \ref{a}) for retarded and advanced
Green functions \cite{KramerPesch,Kopnin-book}
\begin{eqnarray}
\zeta_S ^{R(A)} &=&\frac{\hbar v_{\parallel }e^{-K}}{2Q\left[ \epsilon \pm i\delta -\epsilon _{0}\right]}
\ ,\;\\
\theta_S ^{R(A)} &=& \frac{2}{\hbar v_\parallel}\int_0^s(\epsilon -\Delta _{0}b/\rho )\zeta_S ^{R(A)}ds^{\prime }
\ ,\;
\end{eqnarray}
coincides with (\ref{zeta}-\ref{K}). These expressions hold as
long as $|\zeta_S|$ exceeds $|b|/\rho$. For $s\gg \xi_S$ the
function $\zeta_S$ assumes its asymptotic expression
$\zeta^{R(A)}_S=-b/\rho$ corresponding to the boundary conditions
\eqref{fg-asimpt}.

In  the clean limit  the vortex potentials induced in the 2D layer
 are crucially dependent on tunneling model.
Assuming specular quasiparticle reflection at the superconductor
surface we put  $\check g_S(+p_z)=\check g_S(-p_z)$ in Eq.
(\ref{selfen-coh}) so that the self energy coincides with the
Green function in the bulk for coherent tunneling $\Sigma_1 =
i\Gamma g_S$, $\Sigma_2 = i\Gamma f_S$  and with its values
averaged over the ensemble for the incoherent one: $\Sigma_1 =
i\Gamma \left<g_S\right>$, $\Sigma_2 = i\Gamma \left<f_S\right>$.
The ensemble averaging in terms of quasiclassical Green functions
is equivalent to the averaging over the 3D momentum direction. One
can separate two terms in the Green function expressions:
\begin{equation}
g^{R(A)}=i\zeta^{R(A)} = \wp \frac{i \hbar v_{\parallel
}e^{-K}}{2Q\left[ \epsilon -\epsilon _{0}\right] }\pm  \frac{\pi
\hbar v_{\parallel }e^{-K}}{2Q}\delta(  \epsilon -\epsilon _{0}) \ .
\label{Gfunct/vortcore}
\end{equation}
The first term has to be taken as a principal value integral when
calculating the angular averages. The second term is proportional
to the delta function of energy and determines the density of
states (DOS) of the vortex core states in the bulk SC. Similarly,
the anomalous functions
\begin{eqnarray}
f^{R(A)}=e^{i\phi}(i\zeta^{R(A)}-\theta^{R(A)})[b+is]/\rho\ , \label{Ffunct/vortcore}\\
f^{\dagger R(A)} =e^{-i\phi}(i\zeta^{R(A)}+\theta^{R(A)})[b-is]/\rho\ .  \label{F+funct/vortcore}
\end{eqnarray}
can be separated into the principal value part and the
delta-functional contribution.

Performing averaging over the polar $\theta_p$ and azimuthal
$\alpha$  angles we take into account the symmetry of the
functions under the $s$-inversion transformation. Thus, we find
the following expressions for the self energy terms:
\begin{eqnarray}
\Sigma_1&=& -\Gamma\left<\zeta_S(s)\right>\label{Sigma_1_incoh}\\
\Sigma_2e^{-i\phi} =\Sigma_2^{\dagger} e^{i\phi}
&=&\Gamma\left<\theta_S(s)s-\zeta_S(s)b\right>/{\rho} \ .
\end{eqnarray}

It is convenient to split the off-diagonal induced potential into
the localized ($\Sigma_2^{loc}$) and the long-range parts:
\begin{eqnarray}
\Sigma_2e^{-i\phi}&=& \Gamma\Phi+\Sigma_2^{loc}\label{Sigma_I_incoh}
\ ,\\
\Phi(\rho)&=&
 \wp  \left<\, {I(s){\rm sign} (s)}/{2Q\left[ \epsilon -\epsilon _{0}\right]
}\right> \ .\label{Phi}
\end{eqnarray}
Here we put $I(s)=2\int_{0}^{s}(\epsilon -\Delta _{0}b/\rho
)e^{-K(s^{\prime })}\,ds^{\prime }$. The long-range function
$\Phi$ can be regarded as an adiabatic induced superconducting
gap. Hereafter we focus on the evaluation of the localized part
which is most important in the primary core region. Averaging over
the azimuthal trajectory angle $\alpha$ we find:
\begin{eqnarray*}
\Re \Sigma_2^{loc}
&=&\Gamma\left<\frac{\hbar v_\parallel e^{-K}}{2Q\Omega\rho }
\left[1-\frac{|\epsilon|}{\sqrt{\epsilon^2 -\Omega^2\rho^2}}
\chi(\epsilon^2 -\Omega^2\rho^2)\right]\right>_z \ ,\\
\Im \Sigma_2^{loc}&=&
\pm\Gamma\left< \frac{\epsilon \hbar v_\parallel e^{-K}}
{2Q\rho \Omega \sqrt{\Omega^2\rho^2 -\epsilon^2}}\chi(\Omega^2\rho^2 -\epsilon^2)\right>_z \ ,\\
\Re \Sigma_1
&=&-{\rm sign}(\epsilon) \Gamma \left< \frac{\hbar v_\parallel e^{-K}}
{2Q\sqrt{\epsilon^2-\Omega^2\rho^2}} \chi(\epsilon^2-\Omega^2\rho^2)\right>_z  \ ,\\
\Im \Sigma_1
&=& \pm\Gamma\left<\frac{\hbar v_\parallel e^{-K}}
{2Q\sqrt{\Omega^2\rho^2 -\epsilon^2}}\chi (\Omega^2\rho^2 -\epsilon^2)\right>_z \ .
\end{eqnarray*}
Here the upper (lower) sign corresponds to a retarded (advanced)
self energy term, $\Omega = d\epsilon_0/db$,
$$\chi(x)=\left\{1,\quad x>1\atop 0,\quad x<1\right. $$
is the Heaviside theta-function, and we use the notation
\[
\left< \ldots\right>_z = \frac{1}{2} \int_0^\pi \sin \theta_p d\theta_p\left(\ldots\right)
\]
for the average over the polar angle $\theta_p$ of the 3D Fermi
momentum. Note that our calculations are essentially based on the
first-order approximation in the small parameter $b/\rho$.
According to Eq.~\eqref{Sigma_RI} the symmetrical
$\Sigma_I(-s)=\Sigma_I(s)$ and antisymmetrical
$\Sigma_R(-s)=-\Sigma_R(s)$ parts of the off-diagonal self energy
term $\Sigma_2e^{-i\phi}$ can be rewritten as follows:
$\Sigma_R=\Sigma_2e^{-i\phi}s/\rho$ and
$\Sigma_I=\Sigma_2e^{-i\phi}b/\rho$.
\section{Scale separation inside the multiple vortex core.}\label{app3-scale-separation-solution}

In this Appendix we present the details of the analytical
procedure used to match the solutions of quasiclassical equations
through the primary core region. In a clean 2D layer we start our
consideration of the induced vortex states  from the Eilenberger
equations (\ref{eq-zeta-2}-\ref{eq-g-2})
 for retarded and advanced Green functions.
In the low energy limit $\epsilon\ll\Delta_\infty$ appropriate
boundary conditions far from the induced vortex core ($\rho\gg
\xi_{2D}$) take the form:
\begin{eqnarray}
\theta &=&\frac{\Gamma s/\rho}{\sqrt{\Gamma^2-\epsilon^2}} \ , \;
\zeta =\frac{-\Gamma b/\rho}{\sqrt{\Gamma^2-\epsilon^2}} \ ,
g=\frac{%
-i\epsilon }{\sqrt{\Gamma ^{2}-\epsilon ^{2}}} \
.\label{bcond-Gamma}
\end{eqnarray}
The self energy terms reveal a quite different behavior in the
small ($\rho\lesssim\xi_S$) and large ($\rho\gg\xi_S$) distance
regions. To match the solutions in these domains we introduce a
certain distance $\rho_0$ such that $\xi_S\ll\rho_0\ll\xi_{2D}$
and consider the Green functions in two overlapping spatial
intervals: (i) $\rho<\rho_0\ll\xi_{2D}$ and (ii) $\rho\gg\xi_S$.

Outside the primary  core region $\rho\gg\xi_S$ Eqs.
(\ref{eq-zeta-2}-\ref{eq-g-2}) for both tunneling models and
arbitrary disorder rate inside the superconductor take the form:
\begin{eqnarray}
\hbar V_{2D} \frac{d\zeta}{ds} +2 \epsilon   \theta   -2i g \Gamma s/\rho =0  , \label{new1}\\
\hbar V_{2D} \frac{d\theta }{ds} -2 \epsilon   \zeta  -2i g \Gamma b/\rho=0 , \label{new2}\\
\hbar V_{2D} \frac{d g}{ds} + 2i \theta\Gamma b/\rho +2i\zeta
\Gamma s/\rho =0  \ .\label{new3}
\end{eqnarray}
The functions $g$ and $\zeta$ are even in $s$ while $\theta$ is
odd, so we can consider only positive $s$ values. We obtain the
solution of the above equations using the first order perturbation
theory in the impact parameter $b$: $\check w(s)=\check
w_0(s)+\check w_1(s)$, where $ \check w(s) =\left(\zeta, \theta,
g\right)^T$. This approximation holds for $|b|\ll \xi_{2D}$. The
zero order in $b$ solution reads
\begin{equation}
\check w_0(s) =\frac{1}{\sqrt{\Gamma^2-\epsilon^2}}\check u_0(s)
+\frac{C}{\sqrt{\Gamma^2-\epsilon^2}}\check u_-(s) \ , \label{g2}
\end{equation}
where
\begin{eqnarray*}
\check u_{\pm}(s)= \left( \begin{array}{c}
\sqrt{\Gamma^2-\epsilon^2}\\ \pm \epsilon \\
 \mp i\Gamma\end{array} \right) e^{\pm \lambda s}\ ,\\
 \check u_0(s) = \left(\begin{array}{c} 0\\ \Gamma \\ -i \epsilon \end{array}\right)\ , \,
 \lambda = \frac{2\sqrt{\Gamma^2 -\epsilon^2}}{\hbar V_{2D}}
\end{eqnarray*}
This solution satisfies the boundary conditions  $g= -i\epsilon
/\sqrt{\Gamma^2 -\epsilon^2}$, $\zeta =0$ and $\theta
=\Gamma/\sqrt{\Gamma^2 -\epsilon^2}$ for $s \to \infty$ and
$\epsilon^2 <\Gamma^2$.
 The first order correction $\check w_1$ can be written as
\begin{equation}
\check w_1(s) =\frac{C_0(s)}{\sqrt{\Gamma^2-\epsilon^2}}\check u_0
  +\frac{C_+(s)}{\sqrt{\Gamma^2-\epsilon^2}}\check u_+ +\frac{C_-(s)}{\sqrt{\Gamma^2-\epsilon^2}}\check u_-
  \ , \label{g-corr}
\end{equation}
where
\begin{eqnarray}
\xi_{2D}C_0(s)=2 C b \int_{s}^\infty e^{-\lambda s}\frac{ds}{\rho}\ , \label{C0} \\
\xi_{2D}C_+(s)=- b \int_{s}^\infty e^{-\lambda s}\frac{ds}{\rho}\ , \label{C+}\\
\xi_{2D}C_-(s)=- b \int_{s_{c}}^s e^{\lambda s}\frac{ds}{\rho}\ . \label{C-}
\end{eqnarray}
The lower limit of integration  $s_c$ in $C_-$ has to be taken
$s_c \sim \xi_S$ for trajectories that go through the primary
vortex core, $b\lesssim \xi_S$, so that the logarithmic divergence
is cut off at the distances $\sim \xi_S$ where the long-range
vortex potential $\Phi$ \eqref{Phi} vanishes. For $b \gg \xi_S$ we
have $s_c =0$. The perturbation approach holds as long as $C_0\ll
C$ and $C_+\ll 1$, i.e., as long as $|b|\ll \xi_{2D}$. For
$s\gg\xi_{2D}$ the coefficient $C_0$ decays faster than
exponentially, while
\[
C_+(s)e^{\lambda s} \to C_-(s)e^{-\lambda s} \to -\frac{\Gamma}{2\sqrt{\Gamma^2-\epsilon^2}}\frac{b}{\rho}
\]
such that $\zeta$ is  $
-(b/\rho)\Gamma/\sqrt{\Gamma^2-\epsilon^2}$ and the corrections to
$\theta$ and $g$ vanish as it should be according to
\eqref{bcond-Gamma}. For a small distance $s=s_0$ ($\rho_0^2=s_0^2+b^2$) we have
\begin{eqnarray}
\zeta(s_0)&=&C+ C_+(s_0)+C_-(s_0), \label{zeta2-s0} \\
\theta(s_0)&=&\frac{1}{\sqrt{\Gamma^2-\epsilon^2}}\left[ \Gamma
-\epsilon C+\Gamma C_0(s_0) \right.\nonumber \\
&&+\left. \epsilon [C_+(s_0)-C_-(s_0)] \right], \quad \label{theta2-s0}\\
g(s_0)&=& \frac{i}{\sqrt{\Gamma^2-\epsilon^2}}\left[-\epsilon + \Gamma C
-\epsilon C_0(s_0)\right.\nonumber \\
 &&- \left. \Gamma [C_+(s_0)-C_-(s_0)]\right].\quad \label{g2-s0}
\end{eqnarray}

Consider first trajectories that miss the primary vortex core, i.e., they go at impact parameters $\xi_S\ll b \ll \xi_{2D}$. In this case, the perturbation result Eqs. (\ref{new1}-\ref{new3}) can be applied along the entire trajectory such that one can put $s_0=s_c =0$. The boundary condition for an odd function requires $\theta (0)=0$.
Since in this case $C_-(0)=0$, we find from  Eq. (\ref{theta2-s0})
\[
\Gamma +\epsilon C_+(0)=\epsilon C -\Gamma C_0(0) \ .
\]
Expressing the coefficients $C_0$ and $C_+$  in terms of the
energy $\epsilon_2(b)$ of bound states in the induced vortex core,
$C_0=-2C C_+=C \epsilon_2(b)/\Gamma$, we find
\begin{equation}
C[\epsilon -\epsilon_2(b)]=\Gamma - \epsilon \epsilon_2(b)/2\Gamma
\ ,\label{const2}
\end{equation}
where the energy spectrum $\epsilon_2(b)$ of localized excitations
is given by the Eq.~\eqref{eps2}.  According to Eq. (\ref{const2})
$\epsilon_2(b)$ is the only spectrum branch in the energy interval
$|\epsilon| \ll \Delta_\infty$. The Green function is
\begin{multline}
g(s)=\frac{-i\epsilon}{\sqrt{\Gamma^2-\epsilon^2}}
 + \frac{i\Gamma C}{\sqrt{\Gamma^2-\epsilon^2}}e^{-\lambda s}
-\frac{i\epsilon C_0(s)}{\sqrt{\Gamma^2-\epsilon^2}} \\
-\frac{i\Gamma }
{\sqrt{\Gamma^2-\epsilon^2}}\left[C_+(s)e^{\lambda s} -C_-(s)e^{-\lambda s}\right]
\label{g2-s-highb} \ .
\end{multline}
For $s\gg \xi_{2D}$ we have $C_0 \to 0$, $C_+ e^{\lambda s}-C_-
e^{- \lambda s}\to 0$, so that the first term is the homogeneous
background while the rest terms describe the vortex contribution.
To obtain the retarded function  for $\epsilon^2 >\Gamma^2$ one
has to continue $\sqrt{\Gamma^2-\epsilon^2} $ analytically
throughout the upper half-plane of complex $\epsilon$ keeping $\Re
\sqrt{\Gamma^2 -\epsilon^2}>0$.

The normalized LDOS can be found as a sum over different
trajectories:
\[
N({\bf
r},\epsilon)=\frac{1}{2\pi}\int\limits_0^{2\pi}N_\epsilon(s,b)d\alpha^\prime \ ,
\]
where $s=\rho \cos\alpha^\prime$, $b=-\rho\sin\alpha^\prime$, and
\[
N_\epsilon (s,b)=\frac{1}{2}\left(g^R(s,b)-g^A(s,b)\right) \ .
\]
For $|\epsilon| <\Gamma$, a nonzero LDOS comes only from the
vortex contribution of the second and third terms in
\eqref{g2-s-highb} due to the presence of a pole in the
coefficient $C$ according to Eq. (\ref{const2}).  The Green
functions and LDOS reach their long-distance values
$g=-i\epsilon/\sqrt{\Gamma^2-\epsilon^2}$ and
$N=|\epsilon|/\sqrt{\epsilon^2-\Gamma^2}\chi(\epsilon^2-\Gamma^2)$
as $\rho\to \infty$. For $\rho\gg\xi_S$ the trajectories with
large impact parameters $b\gtrsim\xi_S$ give the main contribution
to the LDOS. In the region ${\xi_S\ll\rho\ll\xi_{2D}}$ we get the
angle--resolved density of states in the form:
\begin{eqnarray}
N_\epsilon(s, b)&=& \frac{\sqrt{\Gamma^2 -\epsilon^2}(\Gamma^2 -\epsilon^2/2)}{\Gamma^2}\nonumber \label{dos<}\\
&&\times \pi \delta[ \epsilon -\epsilon_2(b)]\ , \phantom{223i space} |\epsilon
|<\Gamma \\
N_\epsilon(s, b)&=&\frac{\sqrt{\epsilon^2-\Gamma^2}[\Gamma^2
-\epsilon_2^2(b)/2]}{{\rm sign}(\epsilon)\Gamma^2[\epsilon -\epsilon_2(b)]} \ , \;
|\epsilon |>\Gamma\ . \label{dos>}
\end{eqnarray}
Thus, the corresponding LDOS in the energy interval $|\epsilon|
<\Gamma$ has the only peaks at $\epsilon=\epsilon_2(\pm\rho)$:
\begin{multline}
N(\rho, \epsilon)=
\frac{1}{\pi}\int\limits_{-\rho}^{\rho}N_\epsilon(\sqrt{\rho^2-b^2},
b) \frac{db}{\sqrt{\rho^2-b^2}}=\\= \frac{\sqrt{\Gamma^2
-\epsilon^2}(1 -\epsilon^2/2\Gamma^2)}
{\sqrt{\epsilon_2^2(\rho)-\epsilon^2}}\chi\left[\epsilon_2^2(\rho)-\epsilon^2\right]\
.
\end{multline}
For energies above the induced gap, $|\epsilon|
>\Gamma $, for the same distances the LDOS is monotonically increasing with
$|\epsilon|$
to its normal state value:
\begin{equation}
N(\rho, \epsilon)= \sqrt{\epsilon^2-\Gamma^2}\left[\frac{|\epsilon|}
{2\Gamma^2}+\frac{(1 -\epsilon^2/2\Gamma^2)}{\sqrt{\epsilon^2-\epsilon_2^2(\rho)}}\right]\ .
\end{equation}

The LDOS behavior for small distances $\rho\lesssim\xi_S$ depends
crucially on trajectories with small impact parameters $b$. In this case one has to match Eqs.~(\ref{zeta2-s0})-(\ref{g2-s0}) with the solution obtained in the vortex core region.
For small $s<s_0$ we assume the even parts of the Green function
$g$ and $\zeta$ to be nearly constant, therefore integrating
Eq.~\eqref{eq-theta-2} along the trajectory over $s$ from $0$ to
$s_0$ we find the matching condition for the Green functions:
\begin{equation}
\frac{\hbar V_{2D}}{2}\theta (s_0)=\zeta(s_0)\int_0^{s_0} \Sigma_1\,
 ds + i g(s_0)\int_0^{s_0} \Sigma_I\, ds  \ .\label{bcond-s0}
\end{equation}
This matching
condition determines the
constant $C$. Its poles as a
function of energy and the impact parameter define the eigenstates
of excitations.

While deriving the effective boundary condition \eqref{bcond-s0}
for $b\lesssim\xi_S$, one needs to separate the exponentially
converging parts $\Sigma_{1,I}^{loc}$ at $s\sim\xi_S$ from the
long-distance, $s\gg \xi_S$, asymptotics of $\Sigma_{1,I}$.
For $\epsilon \ll
\Delta_\infty$ the
long-distance expressions, Eq.~\eqref{fg-asimpt} yield $\Sigma_1=0$,
$\Sigma_R =\Gamma s/\rho$, $\Sigma_I = \Gamma b/\rho$. Therefore,
we find
\begin{equation}\label{Sigma_1_int}
\int_0^{s_0} \Sigma_1\, ds \approx \int_0^{\infty} \Sigma_1^{loc}\, ds \ ,
\end{equation}
\begin{multline}\label{Sigma_I_int}
\int_0^{s_0} \Sigma_I\, ds =\int_0^{\xi_S} \Sigma_I^{loc}\,
 ds +\Gamma\int_{\xi_S}^{s_0} b/\rho \, ds \approx \\
 \approx \int_0^{\infty} \Sigma_I^{loc}\, ds +\Gamma b \ln({s_0}/{\xi_S}) \ .
\end{multline}
The localized self-energy  parts $\Sigma_{1,I}^{loc}$ determine
the small-distance LDOS and spectrum of excitations. Therefore,
while $\Sigma_{1,I}^{loc}$ are dependent on the tunneling model,
we should consider these models separately.

\subsection{Coherent Tunneling}\label{app3-subsec-Coh}

Here we consider the quasiparticle trajectories  which go through
the core of the primary vortex at impact parameters $b\ll \xi_S$
assuming coherent tunneling mechanism and derive the expressions
for the spectrum of localized excitations and LDOS. In this case,
the self energies are equal to the quasiclassical Green functions
in the bulk SC taken at the same trajectory as in the 2D layer
(Fig.~\ref{fig-2Dvortex-coords}):
\begin{equation}
\Sigma_R=\Gamma \theta_S\ , \;
\Sigma_I=-\Gamma \zeta_S \ .
\end{equation}

Note that the localized part $\Sigma_2^{loc}$ of the effective
order parameter $\Sigma_2$ has the coordinate dependence
$\Sigma_2^{loc}=i\Sigma_I^{loc}(b,s)e^{i\alpha}$ with \emph{zero}
circulation, unlike its adiabatic part \eqref{fg-asimpt}
$\Sigma_2(\rho\gg\xi_S)=\Gamma e^{i\phi}$. As we will see below it
is this different angular dependence of the effective gap
asymptotics, which leads to the formation of a ``shadow'' of the
bulk SC anomalous branch in the excitation spectrum and LDOS in
the 2D layer.

Eqs. (\ref{C0}--\ref{C-}) yield
\begin{eqnarray}
C_0(s_0) &=& \frac{2 C b }{\xi_{2D}} \ln \frac{1}{\lambda s_0} \ , \\ 
C_+(s_0)\pm C_-(s_0)&\approx& -\frac{b}{\xi_{2D}} \ln
\frac{1}{\lambda\xi_S}\approx-\frac{\epsilon_2(b)}{2\Gamma} \ .
\end{eqnarray}
We now match the asymptotic solution
Eqs.~(\ref{zeta2-s0}-\ref{g2-s0}) obtained for $s\geq s_0$ using
Eq.~(\ref{bcond-s0}) and
Eqs.~(\ref{Sigma_1_int}, \ref{Sigma_I_int}).
As a result,
\begin{multline}
C\left[ \xi_{2D}[\epsilon-\epsilon_2(b)] +
2[\Gamma- \sqrt{\Gamma^2-\epsilon^2}-
\frac{\epsilon \epsilon_2(b)}{\Gamma}]\int_0^\infty \zeta_0\, ds \right] \\
= \left[\xi_{2D}\Gamma +2\epsilon \int_0^\infty \zeta_0\,
 ds - \xi_{2D} \frac{\epsilon \epsilon_2(b)}{2\Gamma}\right.
 \\
\left.  -(\Gamma +\sqrt{\Gamma^2-\epsilon^2})\frac{\epsilon_2(b)}
{\Gamma}\int_0^\infty \zeta_0 \, ds \right]  \ ,\label{const}
\end{multline}
where $\zeta_0(s)$ is the localized part of $\zeta_S$ and
\[
\int_0^\infty \zeta_0\, ds =\frac{\hbar v_\parallel}{2[\epsilon -\epsilon_0(b)]}\ .
\]
Here we put $g=i\zeta_0$ and replace the cutoff parameter in
\eqref{eps2} by $\Lambda=\xi_{2D}/\xi_S$. For $b \gg \xi_S$ the contributions
from  the primary vortex core
proportional to $\int_0^\infty \zeta_0\, ds$ vanish since the
trajectory misses the core, and Eq. (\ref{const}) goes over into Eq.
(\ref{const2}).

For small $b\ll\xi_S$ the Green function has a pole when
\begin{multline}
P(\epsilon,b)=[\epsilon-\epsilon_2(b)][\epsilon -\epsilon_0(b)]
\\+ q \left[\Gamma^2 -
\Gamma \sqrt{\Gamma^2-\epsilon^2}-\epsilon \epsilon_2(b)\right] =0 \label{pole1}
\end{multline}
where $q= v_\parallel/V_{2D}$. This equation coincides with Eq.~(\ref{pole_coh}) within the accuracy of our approximation since $\epsilon_2(b)\ll
\epsilon_0(b)$.
The coefficient $C$ takes the form
\begin{multline}
C=\frac{[\Gamma-\epsilon \epsilon_2(b)/2\Gamma][\epsilon -\epsilon_0(b)]}{P(\epsilon,b)} \\
+\frac{q[\epsilon\Gamma-\epsilon_2(b)(\Gamma+\sqrt{\Gamma^2-\epsilon^2})/2]}{P(\epsilon,b)}
\end{multline}
Equation \eqref{pole1} has two  real-valued branches of
solutions $\epsilon_{1,2}(b)$ in the range $|\epsilon |<\Gamma$
and one complex branch $\epsilon_{1}(b)$ in the range $\Gamma
<|\epsilon|<\Delta_\infty$  for retarded (advanced) Green functions. For $\epsilon \ll \Gamma$,  expanding
Eq. (\ref{pole1}) in $\epsilon /\Gamma$ within the first order
accuracy in $\epsilon_2(b)$ we can write
\begin{eqnarray}
[\epsilon-\epsilon_2(b)][\epsilon -\epsilon_0(b)] +\frac{q  }{2}[\epsilon-\epsilon_2(b)]^2  =0  \label{pole3}
\end{eqnarray}
This equation  has two solutions:
\begin{equation}
\epsilon_1(b) =(1+q/2)^{-1}\epsilon_0(b) \label{E<G2}
\end{equation}
 and $\epsilon_2(b)$.

The angle-resolved DOS for small energies $|\epsilon|\ll\Gamma$
and $\rho\lesssim\xi_S$ reads
\begin{multline}
N_\epsilon(s,b)=\frac{\pi\Gamma q}{2}\delta[\epsilon-\epsilon_1(b)]+\frac{\pi\Gamma(q+2)}{2}\delta[\epsilon-\epsilon_2(b)] \ .
\end{multline}
Here we neglect the terms $\epsilon\epsilon_2(b)/\Gamma^2$ and
$\epsilon_2(b)/\epsilon_1(b)$ and put
$\epsilon_0(b)/\epsilon_1(b)=1+q/2$ according to \eqref{E<G2}.
 In this case  the LDOS
\begin{multline}\label{LDOS_Coh_small_E}
N(\rho,\epsilon)=\frac{\Gamma q\chi
[\epsilon_1^2(\rho)-\epsilon^2]}{2\sqrt{\epsilon_1^2(\rho)-\epsilon^2}}+
\frac{\Gamma (q+2)\chi[\epsilon_2^2(\rho)-\epsilon^2]}{2\sqrt{\epsilon_2^2(\rho)-\epsilon^2}}
\end{multline}
reveals a two-peak structure vs energy at
$\epsilon=\epsilon_{1,2}(\rho)$. For $|\epsilon| \sim \Gamma$, one
can neglect $\epsilon_2(b)$ and obtain:
\begin{eqnarray}
[\epsilon -\epsilon_0(b)]\left[\Gamma +
\sqrt{\Gamma^2-\epsilon^2}\right] + q \Gamma \epsilon =0 \ . \label{pole2}
\end{eqnarray}
For $|\epsilon| >\Gamma$ the dispersion relation is complex valued
and for retarded functions takes the form:
\begin{equation}
\epsilon [\epsilon -\epsilon_0(b)] +
 q\Gamma \left[\Gamma   +i {\rm sign}(\epsilon)
 \sqrt{\epsilon^2- \Gamma^2}\right]  =0  \ .\label{E>G}
\end{equation}
The latter equation describes the resonant states in the 2D vortex
core which decay into the quasiparticle waves propagating in the
2D layer above the induced gap.

Finally, the whole spectrum  structure, shown in
Fig.~\ref{fig-spectrum2}, has  two anomalous branches: (i) one of
them $\epsilon_2(b)$ is completely real-valued and follows the
CdGM spectrum for the superconductor with homogeneous gap
$\Gamma$; (ii) another one is close to the bulk CdGM spectrum, but
has a discontinuity at $\epsilon=\Gamma$, where it becomes
essentially complex.

Thus, the LDOS for energies above the induced gap $|\epsilon|>\Gamma$ and small distances $\rho, b\lesssim\xi_S$ reads
%
\begin{multline}\label{LDOS_Coh_big_E}
N(\rho ,\epsilon) =\frac{\sqrt{\epsilon^2-\Gamma^2}}{|\epsilon|} \\
+ \frac{q\Gamma^2}{2|\epsilon|}\Re\frac{\sqrt{\epsilon^2-\Gamma^2}-i\Gamma}{\sqrt{(\epsilon^2+q\Gamma^2+i q\Gamma\sqrt{\epsilon^2-\Gamma^2})^2-\epsilon^2\epsilon_0^2(\rho)}}
\end{multline}
and has the only peak at $\epsilon=\Re\epsilon_1(\rho)$ of the
height $\sim\Gamma^2/\epsilon_0^2(\rho)$ for
$\rho\gtrsim\xi_S^2/\xi_{2D}$. In the opposite limit of rather
large distances $\rho>\xi_S^2/\xi_{2D}$ at $|\epsilon| > \Gamma$,
the spectrum reduces to the CdGM spectrum with a finite
broadening:
\begin{equation}
\epsilon_1(b) =\epsilon_0(b) -i\Gamma q\label{E>>G} \ .
\end{equation}
The LDOS has a small difference from its normal state value
$N_0=1$:
\begin{equation}
N(\rho ,\epsilon) =1+ \frac{q\Gamma^2}{2\epsilon^2}\Re\frac{|\epsilon|-i\Gamma}{\sqrt{(\epsilon+i q\Gamma)^2-\epsilon_0^2(\rho)}}
\end{equation}

The LDOS in the whole energy range (\ref{LDOS_Coh_small_E},
\ref{LDOS_Coh_big_E}) has two or even three peaks for such
distances. The latter case is realized at the distances
corresponding to $b'<b<b^*$, where the spectrum vs the impact
parameter has 3 anomalous branches.
%

\subsection{Incoherent Tunneling}\label{app3-subsec-Incoh}
 Assuming
small impact parameter values $b\ll\xi_S$, i.e.
$\epsilon_2(b)\ll\Gamma^2/\Delta_\infty$, we obtain an expression
for the coefficient $C$ using the asymptotical solution (\ref{g2},
\ref{g-corr}) and the matching condition \eqref{bcond-s0}:
\begin{multline}
 C\left[ \epsilon -\epsilon_2(b) + \frac{2\sqrt{\Gamma^2-\epsilon^2}}{\hbar V_{2D}} \int_0^{\infty} \Sigma_1\, ds \right. \\
\left. -\frac{2\Gamma}{ \hbar V_{2D}} \int_0^{\infty} \Sigma_I^{loc} \,
ds\right] =\left[ \Gamma -\frac{2\epsilon}{ \hbar V_{2D}}
\int_0^{\infty} \Sigma_I^{loc} \, ds\right] \quad
\label{eqC-incoh}
\end{multline}
Since $|\Sigma_1| \sim
|\Sigma_I^{loc}| \sim \Gamma $ the pole of the coefficient $C$ is located at small energies $\epsilon \lesssim
\Gamma^2/\Delta\ll \Gamma$. Thus, for $\epsilon \ll \Gamma$ the expression for this coefficient takes the form
\begin{eqnarray}
C\left[ \epsilon -\epsilon_2(b) + \frac{2}{ \xi_{2D}}
\int_0^{\infty}\left( \Sigma_1-\Sigma_I^{loc} \right) \, ds\right]
=\Gamma \ . \quad  \label{eqC-incoh-1}
\end{eqnarray}
The localized self energies $\Sigma_1$ and  $\Sigma_I^{loc}$ can be
neglected for $\epsilon \sim \Gamma$. They also vanish for $|b|\gg
\xi_S$. In both these limits, Eq.~\eqref{eqC-incoh} transforms
into Eq.~\eqref{const2}. The integral term in the
 equation above can be written in terms of its real ${\beta(b)=\beta_I(b)-\beta_1(b)}$ and
 imaginary ${\gamma(b)=\gamma_I(b)-\gamma_1(b)}$ parts
as follows:
\begin{equation}\label{beta-gamma-def}
\frac{2}{\xi_{2D}}\int_0^{\infty}\left(
\Sigma_1-\Sigma_I^{loc} \right) \, ds=-\beta(b)\pm
i\gamma(b) \ .
\end{equation}
Here upper (lower) sign corresponds to the retarded (advanced) Green function.
Further we calculate the terms of real $\beta_{1,I}$ and imaginary $\gamma_{1,I}$ parts of the integral \eqref{beta-gamma-def},
 which are defined by the following expressions
\begin{eqnarray*}
\beta_\alpha(b)=\frac{2}{\xi_{2D}}\int\limits_0^\infty\Re\Sigma_\alpha(s)ds
\ , \quad
\gamma_\alpha(b)=\frac{2}{\xi_{2D}}\int\limits_0^\infty\Im\Sigma_\alpha(s)ds
\end{eqnarray*}
and consider the case of the small impact parameter values $b\ll \xi_S$:
\begin{eqnarray*}
\beta_I(b) &=&\frac{2\Gamma^2 b}{V_{2D}}\int_0^\infty \left< \frac{v_\parallel e^{-K}}{2Q\Omega\rho^2 }\right. \\
&&\times \left. \left[
1-\frac{|\epsilon|}{\sqrt{\epsilon^2-\Omega^2\rho^2}}\chi(\epsilon^2-\Omega^2
\rho^2)\right] \right>_z\, ds \ ,
\end{eqnarray*}
where $\rho^2 =b^2+s^2$. In this case the first term in the above
integral is determined by $s \sim b$:
\begin{eqnarray*}
\Gamma b\int_0^\infty \left< \frac{v_\parallel e^{-K}}{Q\Omega\rho^2 }
\right>_z\, ds=
\Gamma b\int_0^\infty \left< \frac{v_\parallel }{Q\Omega(s^2+b^2) } \right>_z\, ds\\
={\rm sign}(b)\Gamma \left< \frac{\pi v_\parallel }{2Q\Omega}
\right>_z \ .
\end{eqnarray*}
The second one is determined by very small impact parameters and
reads:
\[
\int_0^{b_0}\frac{ds}{\sqrt{b_0^2-s^2}}=\frac{\pi}{2}\ , \;
\int_0^{b_0}\frac{ds}{(s^2+b_0^2)\sqrt{b_0^2-s^2}}=\frac{\pi
\Omega}{2|b\epsilon|} \ ,
\]
where $b_0^2=\epsilon^2/\Omega^2 -b^2>0$. As a result, we find:
\begin{eqnarray*}
\beta_I(b) ={\rm sign}(b) \frac{\Gamma^2}{V_{2D}} \left< \frac{\pi
v_\parallel }{Q\Omega} \chi(\Omega^2b^2-\epsilon^2) \right>_z \ ,
\end{eqnarray*}
\begin{eqnarray*}
\beta_1(b) =-{\rm sign}(\epsilon)\frac{\Gamma^2}{V_{2D}}\left<
\frac{\pi v_\parallel }{Q\Omega}\chi(\epsilon^2-\Omega^2b^2)
\right>_z \ . \label{real1}
\end{eqnarray*}
After simplifying the expression for
$\beta(b)=\beta_I(b)-\beta_1(b)$ we obtain \eqref{beta-res}. For
$b\gtrsim \xi_S$ the quantity $\beta(b)$ decays as $\exp
(-2b/\xi_S)$.

The expressions for imaginary parts hold for any distances $\rho$
because the delta functions in the integrals
select only the trajectories that pass at small impact parameters:
\begin{eqnarray*}
\gamma_1(b) =\frac{\Gamma^2}{V_{2D}} \int_0^\infty
\left<\frac{v_\parallel e^{-K}}{Q\sqrt{\Omega^2\rho^2 -\epsilon^2}}\chi (\Omega^2\rho^2 -\epsilon^2)\right>_z\, ds\\
=\frac{\Gamma^2}{V_{2D}} \left<\frac{v_\parallel
}{Q\Omega}\ln\frac{\Delta_\infty}{\sqrt{|\Omega^2b^2-\epsilon^2|}}\right>_z
\ ,
\end{eqnarray*}
\begin{eqnarray*}
\gamma_2(b) =\frac{\Gamma^2 b}{V_{2D}} \int_0^\infty
\left<\frac{\epsilon}{\Omega \rho^2}\frac{v_\parallel e^{-K}}
{Q\sqrt{\Omega^2\rho^2 -\epsilon^2}} \chi (\Omega^2\rho^2 -\epsilon^2)\right>_z\, ds\\
={\rm sign}(b\epsilon)\frac{\Gamma^2}{V_{2D}} \left<\frac{v_\parallel
}{Q\Omega}\ln\frac{\Omega|b|+|\epsilon|}
{\sqrt{|\Omega^2b^2-\epsilon^2|}}\right>_z \ .
\end{eqnarray*}
Here we use the following expressions for the standard integrals:
\begin{eqnarray*}
\int_{b_0}^{s_{max}}\frac{ds}{\sqrt{s^2\pm b_0^2}}=
\ln\frac{\Delta}{\sqrt{|\Omega^2b^2-\epsilon^2|}} \ ,
\end{eqnarray*}
where $s_{max}\sim \xi_S$, and
\begin{eqnarray*}
\int_{b_0}^{s_{max}}\frac{ds}{\sqrt{s^2\pm b_0^2}(s^2+b^2)}=
\frac{\Omega}{|b\epsilon|}\ln\frac{\Omega |b|+ |\epsilon|
}{\sqrt{|\Omega^2b^2-\epsilon^2|}} \ .
\end{eqnarray*}
The imaginary terms also decay exponentially for $b\gtrsim \xi_S$.
The expression for $\gamma(b)=\gamma_1(b)-\gamma_I(b)$ gives
\eqref{gamma-res}. As a result, the expression for the coefficient
$C$ reads:
\begin{eqnarray}
C=\Gamma/\left[ \epsilon -\epsilon_2(b)-\beta(b) +i\gamma(b)\right]
\end{eqnarray}
and the angle-resolved DOS for $\epsilon < \Gamma$ takes the form
\begin{eqnarray}
N_\epsilon(s,b)= \frac{\Gamma \gamma(b)e^{-\lambda|s|}}{[\epsilon-\epsilon_2(b)-
\beta(b)]^2 +\gamma^2(b)}
\end{eqnarray}
 coinciding with Eq.~\eqref{ARDOS} in the main text.
Since parameters $\beta, \gamma \sim \Gamma/\Delta$ and $\epsilon_2(b)/\Gamma \ll1$ are small
 for $b\ll \xi_{2D}$ and $|\epsilon|>\Gamma$, the LDOS reaches its bulk value in this region:
\begin{eqnarray}
N(\rho,\epsilon)=\frac{\sqrt{\epsilon^2-\Gamma^2}}{|\epsilon|} \ .
\label{dos3}
\end{eqnarray}

\subsection{Dirty superconductor and clean 2D Layer}\label{app3-subsec-Dirty}
Here we derive the Green functions and the DOS in the 2D layer for
the dirty limit of the bulk SC. For small impact parameter values
$b\ll \xi_S$ we get $\Sigma_I^{loc}=0$  and  the matching
condition takes the form:
\begin{eqnarray}
\xi_{2D} \theta(s_0) =2i\zeta(s_0) \int\limits_0^{\infty}
\sin\Theta\, ds
+2ig(s_0) b \ln [s_0/\xi_S] \quad \label{bcond-theta2-dirty} \ .
\end{eqnarray}
The coefficient $C$ in this case has the only broadened pole at
$\epsilon=\epsilon_2(b)$:
\begin{eqnarray}
C\left[ \epsilon -\epsilon_2(b) + i\gamma \right] =  \Gamma
\label{eqC-dirty} \ ,
\end{eqnarray}
where the broadening
\[
\gamma = \frac{2\Gamma \sqrt{\Gamma^2-\epsilon^2}}{\hbar V_{2D}}
\int_0^{\infty} \sin\Theta \, ds
\]
coincides with Eq.~\eqref{gamma-res-dirty} in the main text.
For $|\epsilon| < \Gamma$ and $\rho<\xi_S$ 
the angle-resolved DOS can be written in the form
\begin{eqnarray}
N_\epsilon(s,b)= \frac{\Gamma^2 }
{\sqrt{\Gamma^2-\epsilon^2}}\frac{\gamma(b)e^{-\lambda|s|}}{[\epsilon-\epsilon_2(b)]^2
+\gamma^2(b)} \label{DOS-dirty<} \ .
\end{eqnarray}
Consequently, the LDOS has a peak of the height
$\sim\Gamma/\gamma(\rho)$ at energy $\epsilon=\epsilon_2(\rho)$.

For the energies above the induced gap $\epsilon >\Gamma$ and small impact parameter values $\epsilon _2(b),\gamma(b)\ll \Gamma$
the local DOS can be replaced by its bulk value:
\begin{equation}
N (\rho,\epsilon) = \frac{\sqrt{\epsilon^2 -\Gamma^2}}{|\epsilon|}
\label{DOS-dirty>}
\end{equation}

For $b\gg \xi_S$ the imaginary part  of energy decays
exponentially, and Eq. (\ref{eqC-dirty}) transforms into Eq.
(\ref{const2}).

\subsection{Dirty superconductor and dirty 2D layer}\label{app3-subsec-Dirty-all}
At the end of this section we concentrate our attention on the
dirty limit both in 2D layer and superconductor. In this case the
bulk SC \eqref{selfen-dirty} and 2D layer Green functions satisfy
the Usadel equations (\ref{eqUsadel1}, \ref{eqUsadel-2D}). Indeed,
for momentum-orientation-averaged Green functions in 2D layer
$$\check g(\rho)=\begin{pmatrix}g_2 & f_2e^{i\phi}\\ -f_2^\dagger e^{-i\phi} & \bar{g}_2\end{pmatrix}=\int\frac{d^2p}{(2\pi)^2}\check g({\bf p, r})\ $$
one can derive the equation:
\begin{multline}
i D_{2D}\left[g_2(\nabla^2-\rho^{-2})f_2-f_2\nabla^2g_2\right]-\\
-2(\epsilon+\Sigma_1)f_2+2\tilde\Sigma_2g_2=0 \ ,
\end{multline}
with $\tilde\Sigma_2=\Sigma_2e^{-i\phi}$.

Using a standard parametrization $\check
g(\rho)=\tau_3\sin\Psi+\tau_2\cos\Psi e^{-i\tau_3\phi}$ and the
expressions for the vortex potentials one can obtain
\eqref{eqUsadel-2D} from the main text with
$\nabla^2=\rho^{-1}\partial_\rho(\rho\partial_\rho)$.
Integrating Eq.~\eqref{eqUsadel-2D}, multiplied by $\rho$, in a
small region around the origin (from $\rho=0$ to the value
$\xi_S\ll\rho_0\ll\xi_{2D}$) we find the matching condition for
the adiabatic Green function (\ref{g2}, \ref{g-corr}):
\begin{multline}
D_{2D}\left[\left.\rho\frac{\partial}{\partial\rho}\Psi\right|_0^{\rho_0}+\int_0^{\rho_0}\frac{\sin2\Psi}{2\rho}d\rho\right]-\\
-2\int_0^{\rho_0}\rho d\rho\left[\Gamma\sin(\Psi-\Theta)+i\epsilon\cos\Psi\right]=0 \ .
\end{multline}

Considering the expansion $\Psi(\rho_0)=\Psi_0-K\rho_0$ with
$K={\partial\Psi(\rho_0)}/{\partial\rho}\sim\xi_{2D}^{-1}$ and
assuming $\Psi_0\neq\pi/2$ one obtains
$
{\cos\Psi_0\approx{\rho_0^2}/{(\xi_{2D}^2\ln\left({\rho_0}/{\xi_S}\right))}\ll
1} $.
This estimate confirms the conclusion that
the LDOS in the dirty limit follows the bulk LDOS
pattern scaled with the 2D coherence length $\xi_{2D}$ to within the second order terms in the small parameter
$\rho_0/\xi_{2D}$.

\end{document}